\documentclass[aps,onecolumn,floatfix,superscriptaddress]{revtex4}
\usepackage{amssymb}
\usepackage{graphicx}
\usepackage{dcolumn}
\usepackage{amsmath}
\usepackage{bm}
\usepackage{textcomp}
\usepackage{epstopdf}
\usepackage{color}
\usepackage{ulem}
\usepackage[toc,page,title,titletoc,header]{appendix}
\usepackage{setspace}

\newcommand{\be}{\begin{eqnarray}}
\newcommand{\ee}{\end{eqnarray}}

\newcommand{\bfk}{{\bf k}}

\begin{document}

\title{Superfluidity enhanced by spin-flip tunnelling in the presence of a magnetic field}
\author{Jun-Hui Zheng}
\affiliation{Department of Physics, National Tsing Hua University, Hsinchu,
Taiwan}
\author{Daw-Wei Wang}
\affiliation{Department of Physics, National Tsing Hua University, Hsinchu,
Taiwan}
\affiliation{Physics Division, National Center for Theoretical
Sciences, Hsinchu, Taiwan}
\author{Gediminas Juzeli\={u}nas}
\affiliation{Institute of Theoretical Physics and Astronomy, Vilnius
University, A. Go\v{s}tauto 12, Vilnius 01108, Lithuania}

\begin{abstract}
\noindent
It is well-known that when the magnetic field is stronger than a critical value, the spin imbalance can break the Cooper pairs of electrons and hence hinder the superconductivity in a spin-singlet channel. In a bilayer system of ultra-cold Fermi gases, however, we demonstrate that the critical value of the magnetic field at zero temperature can be significantly increased by including a spin-flip tunnelling, which opens a gap in the spin-triplet channel near the Fermi surface and hence reduces the influence of the effective magnetic field on the superfluidity. The phase transition also changes from first order to second order when the tunnelling exceeds a critical value. Considering a realistic experiment, this mechanism can be implemented by applying an intralayer Raman coupling between the spin states with a phase difference between the two layers.
\end{abstract}

\maketitle

\noindent {\bf \large Introduction}\\
Magnetism is generally known to suppress superconductivity when the strength of the magnetic field exceeds a critical value. Survival of superfluidity in the presence of a strong magnetism has been a long-term interesting problem in the condensed matter physics. The central problem is that, in the Bardeen-Cooper-Schrieffer (BCS) theory of superconductivity, electrons form Cooper pairs in the spin singlet channel\cite{Ginzburg1957spj,Berk1966prl}. However, these pairs can be broken if the effective magnetic field is strong enough to flip the spin. This situation applies even if the Cooper pairs are mediated by magnetic fluctuations in some strongly correlated materials \cite{Mathur1998n,Saxena2000n}. A possible exception is probably the theoretical prediction of a so called Fulde-Ferrell-Larkin-Ovchinnikov (FFLO) state \cite{Fulde1964pr,Larkin1965spj}, where the Cooper pair has a finite center-of-mass momentum to form a spatially modulated order parameter \cite{Matsuda2007jpsj,Shimahara2008book,Kenzelmann2008sci,Bianchi2003prl}. Yet, the FFLO states have not yet been experimentally observed neither in condensed matter system  \cite{Matsuda2007j,Beyer2013} nor in the systems of ultracold atoms \cite{Zwierlein2006s,Partridge2006s,Taglieber2008prl}. It is probably because the allowed parameter regime is in general too narrow to be observed.
Another possible coexistence of magnetism and superconductivity arises in scenarios where the Cooper pairs become triplet states through the $p$-wave or $f$-wave interaction due to Pauli's exclusion principle \cite
{Dai2001n,Pfleiderer2001n,Huy2007prl,Machida2001prl,Samokhin2002prb,Nevidomskyy2005prl,Linder2007prb}.

In this paper, we provide a new mechanism to greatly enhance superfluidity of ultracold Fermi gases in a bilayer system with a short range $s$-wave interaction within individual layers. The superfluidity then can survive in a much larger effective magnetic field even without going to the FFLO regime.
This is possible by having a single particle spin-flip tunnelling between the layers. When the tunnelling amplitude exceeds a limiting value, the usual first order phase transition from the superfluid to normal state becomes second order, and the critical value of magnetic field increases almost proportionally to the tunnelling amplitude. Such a behavior can be understood from the fact that the spin-flip tunnelling couples atoms with two different spins in two different layers. This makes the Cooper pairs to include triplet contributions of spins in different layers to fulfil the Pauli exclusion principle.
Similar results can be also observed in a multi-layer structure with a staggered effective magnetic field. Our results may be also relevant to the High $T_c$ superconducting material, where a strong anti-ferromagnetic correlation between nearest-neighboring CuO$_2$ planes is observed through the neutron-scattering experiment \cite{HighTc}.

\begin{figure}[tbp]
\begin{center}
\includegraphics[width=3.32 in]{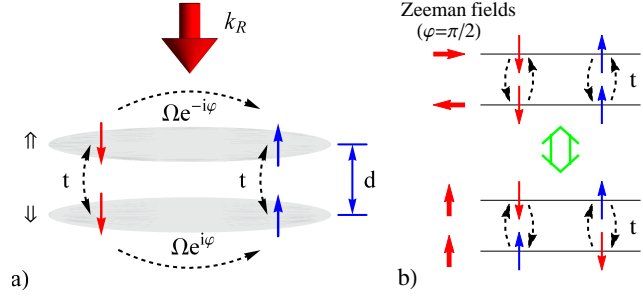}
\end{center}
\caption{ (a) Schematic representation of a bilayer
structure containing two component fermions in individual layers.
 The atoms can undergo spin-independent tunnelling and spin-flip Raman transitions.
The phase difference
$2\protect\varphi =k_{R}d$ of Raman coupling in each layer can be tuned
through  an inter-layer distance $d$  and  a wave-vector of the Raman coupling
$\mathbf{k}_{{\mathrm{R}}}$ oriented perpendicular to the layers. (b)  For $\protect\varphi =\protect\pi /2$,
the  Raman coupling can be represented by an effective Zeeman field antiparallel
in each layer. This
is mathematically equivalent to a
 parallel Zeeman field and a spin-flip tunnelling, as illustrated in a lower part of (b).}
\label{fig1}
\end{figure}

In a realistic experiment, the outlined bi-layer scenario appears to be equivalent to a two-component (spinor)  gas of ultracold atomic fermions loaded into a bi-layer trapping potential with a conventional tunnelling between the layers and a Zeeman magnetic field alternating in different layers, shown in Fig.\,\ref{fig1}. The alternating Zeeman field can be effectively generated by means of a Raman coupling \cite{Deutsch1998,Lin2011,Goldman2014}  within individual layers with a properly chosen  out of plane Raman recoil. The latter recoil provides the phase difference $2\varphi$ of the coupling amplitude in different layers needed for creating the alternating Zeeman field, as depicted in Fig.\,\ref{fig1}a. For $\varphi=\pi/2$ the scheme is mathematically equivalent to a setup involving  a parallel Zeeman field and a spin-flip tunnelling (see Fig.\,\ref{fig1}b).

~\\
\noindent{\bf \large System and Methods}\\
\noindent{\bf System Hamiltonian in original basis}\\
We consider a spin-1/2 Fermi gas trapped in a bilayer potential. In each layer the Raman beams induce spin-flip transitions with Rabi frequencies
\begin{equation}
\Omega _{\pm }=\Omega e^{\pm i\varphi }=\Omega _{x}\pm i\Omega _{y}\quad \,
\label{Omega}
\end{equation}%
where the upper (lower) sign in $\pm $ corresponds to down (up) layer, with $\Omega _{x}=\Omega \cos \varphi $ and  $\Omega _{y}=\Omega \sin \varphi $. The phase difference for the Raman coupling in different layers $2 \varphi \equiv |\mathbf{k}_{{\mathrm{R}}}|d$ is achieved by taking a wave-vector of the Raman coupling ${\mathbf{k}}_{{\mathrm{R}}}$ perpendicular to the layers separated by a distance $d$ (see Fig.\thinspace \ref{fig1}a). The Pauli matrixes for the spin $1/2$ atoms are denoted by $\sigma _{x,y,z}$. On the other hand, it is convenient to treat the layer index as a pseudospin to be represented by the Pauli matrices $\tau _{x,y,z}$.

As a result, the second quantized single particle Hamiltonian describing intralayer Raman transitions and interlayer tunneling can be written as
\begin{equation}
H_{0}=\sum_{\mathbf{k}}\tilde\Psi _{\mathbf{k}}^{\dag }\left[ \varepsilon _{
\mathbf{k}}-t\tau _{x}\otimes \sigma _{0}+\Omega _{x}\tau _{0}\otimes \sigma
_{x}+\Omega _{y}\tau _{z}\otimes \sigma _{y}\right] \tilde\Psi _{\mathbf{k}},
\label{hs}
\end{equation}%
where $\sigma _{0}$ and $\tau _{0}$ are identity matrixes, $\varepsilon _{\mathbf{k}}=\mathbf{k}^{2}/2m-\mu $ measures the kinetic energy with respect
to the chemical potential $\mu $, and $t$ is the interlayer tunneling amplitude. The four component vector field operator $\tilde\Psi _{\mathbf{k}}=[\tilde\psi _{\Uparrow \uparrow ,\mathbf{k}},\tilde\psi _{\Uparrow \downarrow ,\mathbf{k}},\tilde\psi _{\Downarrow \uparrow ,\mathbf{k}},\tilde\psi _{\Downarrow \downarrow , \mathbf{k}}]^{T}$ featured in Eq.~(\ref{hs}) is a column matrix composed of operators $\tilde\psi _{j\gamma ,\mathbf{k}}$ annihilating an atom with a spin $\gamma =\uparrow ,\downarrow $ and a momentum $\mathbf{k}$ in a layer $j=\Uparrow ,\Downarrow $, whereas $\tilde\Psi _{\mathbf{k}}^{\dag }$ is the corresponding raw matrix composed of the creation operators. For brevity in the following, we will omit the identity matrices $\tau _{0}$ and $\sigma_{0}$ in tensor products like $\tau _{0}\otimes \sigma _{x}\equiv \sigma_{x} $ and $\tau _{x}\otimes \sigma _{0}\equiv \tau _{x}$.

The Hamiltonian~(\ref{hs}) describes a quantum system of four combined layer--spin atomic states $|j=\Uparrow ,\Downarrow ;\gamma =\uparrow,\downarrow >$ coupled in a cyclic way (see Fig.\,\ref{fig1}a): $|\Downarrow \downarrow >\rightarrow|\Downarrow \uparrow >\rightarrow |\Uparrow \uparrow >\rightarrow
|\Uparrow \downarrow >\rightarrow |\Downarrow \downarrow >$. The phase $2\varphi $ accumulated during such a cyclic transition allows to control the single particle spectrum \cite{Campbell2011}. The choice of the phase $2\varphi $ affects significantly also the many-body properties of the system, as we shall see later on.

We are considering a short range interaction between the atoms with opposite spins in the same layer. It is described by the following interaction Hamiltonian
\begin{equation}
H_{\mathrm{int}}=g\sum_{j=\Uparrow ,\Downarrow }\int d^{2}\mathbf{r}\tilde\psi
_{j\uparrow }^{\dagger }(\mathbf{r})\tilde\psi _{j\downarrow }^{\dagger }(\mathbf{r%
})\tilde\psi _{j\downarrow }(\mathbf{r})\tilde\psi _{j\uparrow }(\mathbf{r}),
\label{hint}
\end{equation}
where $g$ is the coupling strength. Note that $H_{\mathrm{int}}$ has a symmetry group: $\mathcal{S}\equiv U(2)\times U(2)\times Z_{2}$, where the two $U(2)$s describe the spin rotations in the first and second layer respectively, and $Z_{2}$ is the transpose transformation in pseudospin (layer) space, i.e., $\tilde\psi _{\Uparrow \gamma }\leftrightarrow \tilde\psi _{\Downarrow\gamma }$.

~\\
\noindent{\bf Equivalent description in a rotated basis}\\
The last two terms of Eq.\,(\ref{hs}) represent effective coupling of the spin $\boldsymbol{\sigma }$ with a parallel Zeeman field $\Omega _{x}$ along the $x$-axis and an antiparallel Zeeman field $\Omega _{y}$ along the $y$-axis for the two layers. In order to have a better understanding of the following calculation results, it is convenient to represent the system in another basis. We first apply a unitary transformation
\begin{equation}
U_{\varphi }=\exp \left[ i\frac{\varphi }{2}\tau _{z}\otimes \sigma _{z}\right] \in \mathcal{S}  \label{eq:U_varphi}
\end{equation}%
rotating the spin $\boldsymbol{\sigma }$ around the $z$ axis by the angle $\mp \varphi $ for the up (down) layer. The resulting Zeeman field then
becomes aligned along the $x$-axis in both layers. A subsequent spin rotation $W=\exp {\left[ i\frac{\pi }{4}\sigma _{y}\right] }\in \mathcal{S}$ around
the $y$ axis by the angle $-\pi /2$ transforms $\sigma _{x}$ to $\sigma _{z}$. After the two consecutive transformations the single particle Hamiltonian
takes the form
\begin{equation}
H_{0}=\sum_{\mathbf{k}}{\Psi}_{\mathbf{k}}^{\dag }\left[ \varepsilon _{
\mathbf{k}}-t\cos \varphi \,\tau _{x}-t\sin \varphi \,\tau _{y}\otimes
\sigma _{x}+\Omega \,\sigma _{z}\right] {\Psi}_{\mathbf{k}}.
\label{hsnew}
\end{equation}
The transformed four component field operator ${\Psi}_{\mathbf{k}}=WU_{\varphi }\tilde\Psi _{\mathbf{k}}\equiv \lbrack {\psi}_{\Uparrow
\uparrow ,\mathbf{k}},{\psi}_{\Uparrow \downarrow ,\mathbf{k}},{\psi}_{\Downarrow \uparrow ,\mathbf{k}},{\psi}_{\Downarrow \downarrow ,\mathbf{k}}]^{T}$  is made of components ${\psi}_{j\uparrow ,\mathbf{k}}$ and ${\psi}_{j\downarrow ,\mathbf{k}}$, which are superpositions of the original spin up and down field operators $\tilde\psi _{j\uparrow ,\mathbf{k}}$ and $\tilde\psi_{j\downarrow ,\mathbf{k}}$ belonging to the same layer $j=\Uparrow,\Downarrow $. Note that going to the new basis the spins are rotated differently in different layers.

The transformed single particle Hamiltonian (\ref{hsnew}) corresponds to a bilayer system subjected to a parallel Zeeman field along the $z$-axis for both
layers, with the interlayer tunneling becoming spin-dependent for $\sin \varphi\neq 0$. For $\protect\varphi =\protect\pi /2$ the transformed Hamiltonian
describes a completely spin-flip tunnelling, as illustrated in Fig.\,\ref{fig1}b. The interaction Hamiltonian  given by Eq.\,(\ref{hint}) is invariant
under the transformation $WU_{\varphi }\in \mathcal{S}$, which involves spin rotation within individual layers and thus does not
change the form of $H_{\mathrm{int}}$.

~\\
\noindent{\bf Single-particle spectrum}\\
The single-particle Hamiltonian $H_0$ given by Eq.\,(\ref{hsnew}) can be reduced to a diagonal form via a unitary transformation $V_{\varphi}$ for the field operator ${\Psi}_{\mathbf{k}}$,
\begin{eqnarray}
H_{0} &=&\sum_{j\gamma ,\mathbf{k}}c_{j\gamma ,\mathbf{k}}^{\dagger }\xi
_{j\gamma ,\mathbf{k}}^{\varphi }c_{j\gamma ,\mathbf{k}}\equiv C_{\mathbf{%
k}}^{\dagger }\mathbf{{\Xi }_{k}^{\varphi }}C_{\mathbf{k}}, \\
C_{\mathbf{k}} &=&V_{\varphi }{\Psi}_{\mathbf{k}}\equiv \left[
c_{\Uparrow \uparrow ,\mathbf{k}},\,c_{\Uparrow \downarrow ,\mathbf{k}%
},\,c_{\Downarrow \uparrow ,\mathbf{k}},\,c_{\Downarrow \downarrow ,\mathbf{k%
}}\right] ^{T}.  \label{h_0-diag}
\end{eqnarray}%
where $c_{j\gamma,\mathbf{k}}$  is an annihilation operator for a normal mode characterized by the eigen-energy $\xi^{\varphi}_{j\gamma,\mathbf{k}}$,  with $j=\Uparrow,\Downarrow$ and $\gamma=\uparrow,\downarrow$. Here  $\mathbf{{\Xi}_k^\varphi}$ is a $4\times 4$ diagonal matrix of eigen-energies $\xi^{\varphi}_{j\gamma,\mathbf{k}}$.

In the following we shall concentrate on two specific cases of interest.  (1) In the first case one has $\varphi=0$,  so that $\Omega_y=0$ and $\Omega_x=\Omega$. (2) In the second case the relative phase is $\varphi=\pi/2$, giving $\Omega_x=0$  and $\Omega_y=\Omega$. The first case corresponds to a
spin-independent tunneling and non-staggered Zeeman field (in the transformed representation, Eq.\,(\ref{hsnew})). The second case corresponds to the spin-flip
tunneling and non-staggered Zeeman field along the $z$-axis, as shown in Fig.\,\ref{fig1}b. For these two cases the unitary transformation $V_{\varphi}$ diagonalizing the  single particle Hamiltonian $H_0$ and the corresponding diagonal operator  $\mathbf{{\Xi}_k^\varphi}$ of eigenenergies $\xi^{\varphi}_{j\gamma,\mathbf{k}}$ read:
\begin{eqnarray}
&& V_0=\exp[-i\frac{\pi}{4}\tau_y],\quad
V_{{\pi}/{2}}=\exp[-i\frac{\theta}{2}\tau _{y}\otimes\sigma_{y}]\,, \\
&& \Xi^{0}_{\mathbf{k}}=\varepsilon_\mathbf{k} + t \tau_z + \Omega \sigma_z \,, \quad
\Xi^{\pi/2}_{\mathbf{k}}=\varepsilon_\mathbf{k} + \Omega_t \sigma_z \,, \label{eq6}
\end{eqnarray}
where $\Omega_t=\sqrt{\Omega^2+t^2}$ and  $\exp[i\theta]\equiv \left( \Omega+i t \right) / \Omega_t $.

For $\varphi=0$ the tunneling and Raman coupling  are decoupled in the single particle Hamiltonian (\ref{hsnew}) or (\ref{hs}), so $\Omega$ and $t$ are separable in single particle dispersion $\xi^{0}_{j\gamma,\mathbf{k}}$.  On the other hand, for $\varphi=\pi/2$ there is a term $\tau _{y}\otimes \sigma _{x}$ in Eq.(\ref{hsnew})  which  mixes the interlayer tunneling $t$ and the Raman coupling $\Omega$, so the single particle dispersion $\xi^{\pi/2}_{j\gamma,\mathbf{k}}$  becomes non-separable.  The latter case corresponds to a ring coupling scheme between four atomic states with an overall phase $2\varphi=\pi$ \cite{Campbell2011}. In such a situation the single particle eigenvalues $\xi^{\pi/2}_{j\gamma,\mathbf{k}}=\varepsilon_\mathbf{k} \pm \Omega_t $ are twice degenerate with resect to the index $j=\Uparrow,\Downarrow$. This leads to significant differences in the BCS pairing for the two cases where  $\varphi=0$ and  $\varphi=\pi/2$.

Without including the interaction effects, the chemical potential (Fermi energy) $\mu_F^\varphi$ satisfies $\sum_{j,\gamma} \frac{A}{(2\pi)^2}\int d^2\bfk \Theta[-\xi^\varphi_{j\gamma}]=N $, where $N$ is the total number of particles, $A$ is the area of system and $\Theta$ is a unit step function. We use $\mu_0=\mu_F^0(t=0,\Omega=0,N)$ to represent the chemical potential for noninteracting particles  at $\Omega =t=0$, with $k_{F}^{0}$ being the corresponding Fermi momentum.

~\\
\noindent{\bf General framework in meanfield theory}\\
In the present paper, we are interested in the effects due to attractive interaction between the atomic fermions ($g<0$) in the bilayer system. As usual, a superfluid order parameter can be expected between fermions with opposite spins in the same layer, i.e., $\Delta_{j}=g\left\langle {\psi}_{j\downarrow}(\mathbf{r}){\psi}_{j\uparrow }(\mathbf{r})\right\rangle$, were $\langle \cdots \rangle$ denotes the ground state expectation value. Without a loss of generality we can apply a $U(1)$ transformation ${\psi}_{j\gamma}\rightarrow e^{i\alpha}{\psi}_{j\gamma}$ to make the order
parameter complex conjugated in different layers $\Delta _{\Downarrow}=\Delta _{\Uparrow}^{\ast }\equiv \Delta _{\mathrm{R}}+i\Delta _{\mathrm{I}}$. In general, there may be a phase difference between the order parameters in different layers described by  $\Delta _{\mathrm{I}}$.  As it will be
shown later, the imaginary part $\Delta _{\mathrm{I}}$ is zero for all cases to be considered.

Adopting the BCS mean-field approximation \cite{Chen2012}, the interaction Hamiltonian (\ref{hint}) reduces to the following quadratic form of creation and annihilation field operators in the momentum space:
\begin{equation}
H_{\mathrm{int}}=\frac{2\Delta ^{2}A}{\left\vert g\right\vert }+\sum_{
\mathbf{k},j= \Uparrow , \Downarrow}\left( \Delta_j {\psi}_{j\uparrow ,\mathbf{k}}^{\dagger}{\psi}_{j\downarrow
,-\mathbf{k}}^{\dagger }+H.c.\right) \,,  \label{hint-mf-k}
\end{equation}
where $\Delta =|\Delta _{\Uparrow }|=|\Delta _{\Downarrow }|=\sqrt{\Delta_{\mathrm{R}}^2+\Delta_{\mathrm{I}}^2}$. Consequently we can express the total Hamiltonian, $H=H_{0}+H_{\mathrm{int}}$, in the BCS form in terms of a set of normal single-particle operators $C_{\mathbf{k}}$ and $C_{\mathbf{-k}}^{\dag }$ given by Eq.\,(\ref{h_0-diag}):
\begin{equation}
H =\frac{1}{2}\sum_{\mathbf{k}}[C_{\mathbf{k}}^{\dag },C_{\mathbf{-k}}^T]%
\left[
\begin{array}{cc}
\mathbf{{\Xi }_{k}^{\varphi }} & D_{\varphi } \\
D_{\varphi }^{\dagger } & -\mathbf{{\Xi }_{k}^{\varphi }}%
\end{array}%
\right] \left[
\begin{array}{c}
C_{\mathbf{k}} \\
C_{\mathbf{-k}}^{\dag T}%
\end{array}%
\right]  +2\sum_{\mathbf{k}}\varepsilon _{\mathbf{k}}+\frac{2\Delta ^{2}A}{%
\left\vert g\right\vert },  \label{k0freeHami}
\end{equation}
where
\begin{equation}
D_{\varphi }=V_{\varphi }\left( i\Delta _{\mathrm{R}}\tau _{0}\otimes \sigma
_{y}+\Delta _{\mathrm{I}}\tau _{z}\otimes \sigma _{y}\right) V_{\varphi
}^{T}\,  \label{eq:D_varphi-explicit}
\end{equation}%
describes the mean-field atom-atom interaction responsible for the BCS pairing, and $V_{\varphi }^{T}$ is a transposed diagonalization matrix (for a detailed derivation, see the Section I of the Supplementary Material \cite{Junhui2016}). For$\varphi =0$ and $\pi /2$, we have $D_{0}=i\Delta _{\mathrm{R}}\sigma_{y}+\Delta _{\mathrm{I}}\tau _{x}\otimes\sigma _{y} $ and $D_{\pi/2}=i\Delta_{\mathrm{R}}\cos \theta \sigma _{y}+\Delta _{\mathrm{R}}\sin\theta \tau _{y}+\Delta _{\mathrm{I}} \tau _{z}\otimes\sigma _{y}$, respectively. Denoting $E_{\alpha ,\mathbf{k}}^{\varphi }\geq 0$ (with $\alpha=1,2,3,4$) to be eigenvalues of the first term in Eq.~(\ref{k0freeHami}), representing the Bogoliubov-DeGuinne (BdG) term, one arrives at the following total ground-state energy (see the Section II of the Supplementary Material \cite{Junhui2016}),
\begin{equation}
\mathcal{E}=\frac{2\Delta ^{2}A}{\left\vert g\right\vert }+\sum_{\mathbf{k}}%
\Big[ 2\varepsilon _{\mathbf{k}}-\frac{1}{2}\sum_{\alpha }E_{\alpha ,%
\mathbf{k}}^{^{\varphi }}\Big] .  \label{k0groundenergy}
\end{equation}

Finally, in a 2D Fermi gas, the fluctuation correction for the effective short-range interaction can be accounted by replacing $g^{-1}=-\frac{1}{A}\sum_{\mathbf{k}}1/(\mathbf{k}^{2}/m+\epsilon _{b})$, where $\epsilon _{b}$ is the two-body binding energy \cite{Randeria1989prl,Randeria1990prb,Loktev2001prs}. The superfluid gap equations are then determined by minimizing the total energy, i.e. $\partial
\mathcal{E}/{\partial \Delta _{\mathrm{R/I}}}=0$. On the other hand, the equation $N=-{\partial \mathcal{E}}/{\partial \mu }$, relates the atomic
number $N$ to the chemical potential $\mu $.

Our major aim is to study effects of the interlayer spin-flip tunneling ($\varphi = \pi/2$) on the superfluid properties for the bilayer system in the presence of magnetic field. We will not consider a possible FFLO phase  that results from a mismatch in the Fermi energies for the two spins leading to the finite center-of-mass momentum for the Cooper pairs \cite{Fulde1964pr,Larkin1965spj}. Including the spin mismatch should not substantially affect our major results, since the FFLO regions are in most cases too small to be observed \cite{Partridge2006s,Zwierlein2006s,Matsuda2007jpsj,Youichi2008jpsj,Zheng2014SR}.

~\\
\noindent{\bf \large Results and Discussion}\\
\noindent{\bf  Single layer limit}\\
To better understand results for our bilayer system, it is instructive to consider first a familiar single layer limit \cite{Randeria1989prl,Randeria1990prb,Loktev2001prs} corresponding to zero interlayer tunneling ($t=0$) in the present model. This will allow one to see how the superfluidity is affected by the effective magnetic field provided by the Raman coupling $\Omega$.

\begin{figure}[tbp]
\includegraphics[width=2.6 in]{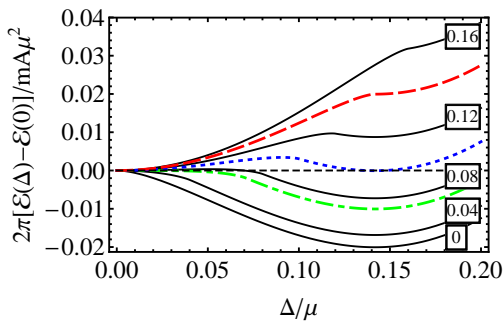}
\caption{(Color online) The ground energy with respect to $\Delta$ for $t=0$.
 The solid (Black) lines correspond to $\Omega=(0
,~0.04,~0.08,~0.12,~0.16) \protect\mu$.
When $\Omega$ is increased to  $\sim \frac{1}{10\protect\sqrt{2}} \protect\mu$
 (dotdashed/Green line), a metastable normal state appears in addition to the superfluid state (Regime II).
When $\Omega_x$ increases to $\frac{1}{10}\protect\mu$ (dotted/Blue line),
the superfluid state becomes metastable (Regime III). Finally, when $\Omega_x $ reaches
$\sim \frac{\protect\sqrt{2}}{10}\protect\mu$ (dashed/Red line), the
metastable superfluid state disappears and the system enters the normal state (Regime IV). We use the parameter
$\protect\epsilon_{b}=0.01\protect\mu_0$ in all calculations.}
\label{fig3}
\end{figure}

For $t=0$ the positive eigenvalues of the BdG operator $E_{\alpha ,\mathbf{k}}^{\varphi }=|\sqrt{\varepsilon _{\mathbf{k}}^{2}+\Delta ^{2}}\pm \Omega |$ exhibit a two-fold degeneracy corresponding to different layers and are independent of $\varphi $ as expected. 
The ground state energy Eq.(\ref{k0groundenergy}) can then be calculated analytically to be (see the Section IIA of the Supplementary Material \cite{Junhui2016}):
\begin{equation}\label{E_tot_t=0}
  \frac{2\pi\mathcal{E}(\Delta)}{mA}= f(\Delta,\epsilon _{b},\mu)
+ \Theta(\Omega-\Delta)g(\Delta,\Omega),
\end{equation}
where $ f(\Delta,\epsilon _{b},\mu)=\Delta ^{2} \ln \frac{\sqrt{\mu ^{2}+\Delta^{2}}-\mu}{\epsilon _{b}} -\frac{\Delta ^{2}}{2}- \mu \sqrt{\mu ^{2}+\Delta ^{2}}-\mu^2$  is the ground-state energy for $\Omega =0$ \cite{Randeria1989prl,Randeria1990prb,Loktev2001prs}. The last term $g(\Delta,\Omega)=\Delta ^{2}\ln \frac{\Omega+\sqrt{\Omega^{2}-\Delta ^{2}}}{\Omega-\sqrt{\Omega^{2}-\Delta ^{2}}}- 2\Omega\sqrt{\Omega^{2}-\Delta ^{2}}$, however, results solely from the finite effective magnetic field, $\Omega $, and comes into play only when $\Omega >\Delta $.  Therefore the superfluidity can not be affected by a relatively small effective magnetic field field acting on the singlet Cooper pairs. In Fig. \ref{fig3}, we show how the ground state energy changes as a function of the
order parameter, $\Delta $, for various values of $\Omega $. We take $\epsilon_{b}=0.01\mu_0 $  in this and the subsequent calculations.

One can determine several important regimes for $\Omega $ where the superfluid order parameter, $\Delta $, can be analytically determined by looking for the global minimum of $\mathcal{E}=\mathcal{E}(\Delta )$:

Regime I corresponds to a limit of small Raman coupling (small effective magnetic field) $0<\Omega <\sqrt{\mu \epsilon _{b}/2}$. In this limit, we have $\Delta =\sqrt{2\mu \epsilon _{b}}>\Omega $. In other words, the last term of Eq. (\ref{E_tot_t=0}) is effectively zero and therefore the superfluid properties are identical to those for a usual 2D BCS state.

Regime II appears for $\sqrt{\mu \epsilon _{b}/2}<\Omega <\sqrt{\mu \epsilon_{b}}$: The obtained superfluid order parameter is still the same, $\Delta =\sqrt{2\mu \epsilon _{b}}>\Omega $. However, the normal state with $\Delta=0 $ becomes meta-stable, i.e., $\partial ^{2}\mathcal{E}/\partial \Delta ^{2}\rfloor _{\Delta =0}=\frac{m A}{\pi} \ln\frac{\Omega^2}{\mu\epsilon_b/2}>0$. In other words, the superfluid state starts to compete in energy with the normal state as the effective magnetic field is increased. Note that $\mathcal{E}(\Delta=0)=-\frac{m A}{\pi}(\Omega^2+\mu^2)$ and $\mathcal{E}(\Delta=\sqrt{2\mu \epsilon _{b}})\simeq -\frac{m A}{\pi}(\mu\epsilon_b+\mu^2$).

Regime III is formed for $\sqrt{\mu \epsilon _{b}}<\Omega <\sqrt{2\mu\epsilon _{b}}$. In that case the last term of Eq.\thinspace (\ref{E_tot_t=0}) is  relevant. The obtained ground state corresponds to the normal state with $\Delta =0$. The superfluid state becomes then a meta-stable state with a finite stiffness.

Regime IV is reached for $\Omega >\sqrt{2\mu \epsilon _{b}}$. In that case the meta-stable superfluid state disappears, and therefore the system transforms to the completely normal state.

We note that the true first order phase transition occurs at the border between the Regimes  II and  III for $\Omega =\sqrt{\mu \epsilon _{b}}$. Yet the appearance of the meta-stable state in the Regimes II  and III effectively broadens the phase transition making it not easily measurable. As we will see later,
the inter-layer tunneling can completely change the situation.

~\\
\noindent{\bf  Zero Raman coupling limit}\\
Next let us suppose there is a non-zero  inter-layer tunneling $t$ and no Raman coupling $\Omega=0$)  \cite{Liu1993prb,Biagini1996prb,Tachiki1990}, so there is no effective magnetic field. In that case one arrives at spin-degenerate eigenvalues of the BdG operator:
$E_{\alpha,\mathbf{k}} ^{\Omega=0}\equiv\sqrt{\varepsilon_{\mathbf{k}}^{2}+\Delta ^{2}+t^{2}\pm 2t\sqrt{\varepsilon_{\mathbf{k}}^{2}+\Delta_{\mathrm{I}}^{2}}}$.
By taking ${\partial \mathcal{E}(\Delta, \Delta_{\mathrm{I}})}/{\partial\Delta_I}= -\frac{1}{2}\sum_{\alpha,\mathbf{k}}{\partial E_{\alpha,\mathbf{k}} }/{\partial\Delta_I}=0$, one gets $\Delta_I=0$.  Thus one finds the following equation for the  ground state energy (see the Section IIB of the Supplementary Material \cite{Junhui2016}):
\begin{equation}  \label{energy23}
\frac {2 \pi}{m A }\mathcal{E}^{\Omega=0}(\Delta) = \frac{1}{2}f(\Delta,\epsilon _{b},\mu+t)+\frac{1}{2}f(\Delta,\epsilon _{b},\mu-t)
\end{equation}
A gap equation is obtained by  taking ${\partial \mathcal{E}}/{\partial\Delta}=0$, i.e.,
\begin{equation}  \label{gapequ2}
\Big[\sqrt{\Delta^2+(\mu-t)^2}-\mu+t\Big]\Big[\sqrt{\Delta^2+(\mu+t)^2}
-\mu-t\Big]=\epsilon_b^2.
\end{equation}

For $\mu-t\gg\Delta$, Eq.(\ref{gapequ2}) yields an asymptotic solution $\Delta=\sqrt{2\epsilon_b\mu_t}$ with $\mu_t=\sqrt {\mu^2-t^2}$, which
goes to the known single layer result, $\Delta=\sqrt{2\epsilon_b \mu}$ \cite{Randeria1989prl}, in the zero tunneling limit ($t\rightarrow 0$). Note that the single particle spectrum is $\xi_{j\gamma ,\mathbf{k}}^{\varphi }=\frac{\bfk^2}{2m}-\mu+(-1)^j t$, so that $\mu-t\gg\Delta$ implies that both of the two bands are occupied. In the other limit, $t-\mu\gg\Delta$, only the states from the lower band could be occupied at zero temperature, and we have $\Delta=\epsilon_b \sqrt{(t+\mu)/{(t-\mu)}}$.

\begin{figure}[tbp]
\includegraphics[width=2.6 in]{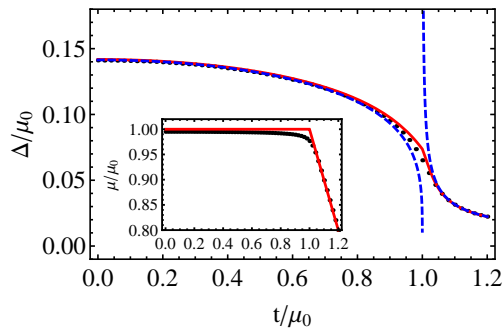}
\caption{ (Color online) The order parameter and chemical potential (inset)
with respect to tunneling strength $t$ for $\Omega=0$. The dotted (Black)
line is obtained by solving numerically the coupled gap equation and the particle number
equation $N=-{\partial \mathcal{E}}/{\partial \mu }$.
 The solid (Red) line corresponds to approximating
$\mu=\mu_F^{\pi/2}$,  where $\mu_F^{\pi/2}$ is the Fermi energy without including the interaction effects. The dashed
(Blue) lines are asymptotic solutions. For inset, the solid (Red) line represents $\mu_F^{\pi/2}$ and
 the dotted (Black) line is a self-consistent numerical result for $\mu$.}
\label{fig2}
\end{figure}

Figure\,\ref{fig2}  shows  a behavior of order parameter as a function of the tunneling strength $t$ for $\Omega =0$. Obviously, the superfluidity decreases with an increase of the tunneling strength, because the inter-layer tunneling plays a role of an effective Zeeman field in the pseudo-spin
(layer) space. Yet now the order parameter decays in a power law in the limit of larger $t$, whereas in the previous case it goes  abruptly to zero with increasing the Zeeman field $\Omega$.

~\\
\noindent{\bf Raman coupling with $\varphi=0$}\\
Now let us consider a  more general case with a finite Raman coupling and a finite interlayer tunneling for $\varphi=0$. In such a situation eigenvalues of the BdG operator have no degeneracy, $E^{\varphi=0}_{\alpha,\mathbf{k}}=\vert E^{\Omega=0}_{\alpha ,\mathbf{k}}\pm \Omega \vert$, with
$E^{\Omega=0}_{\alpha ,\mathbf{k}}\equiv E^{\Omega=0}_{j ,\mathbf{k}}=\sqrt{\Delta_{\mathrm{R}}^{2}+\big[t +(-1)^j \sqrt{\varepsilon_{\mathbf{k}}^{2}+\Delta_{\mathrm{I}}^{2}}~\big]^2} $, where the four values of $\alpha = \{j,\pm\}$ are obtained by combining two values of $j=1,2$ and two values of $\pm$. For the superfluid phase, one should have $E^{\Omega=0}_{\alpha ,\mathbf{k}} \pm \Omega >0 $ for all of $\bfk$  in order to open a gap at the Fermi surface. (In fact,  $E^{\varphi=0}_{\alpha,\mathbf{k}}\pm \Omega$ is a continuous function of the momentum $\bfk$, and goes to $+\infty$ in the limit of large $\bfk$.  If for some $\bfk$  the function becomes negative, it must cross the zero continuously. In such a situation the BdG spectrum will not open the gap at the Fermi surface.) This implies that $\left|\Delta_{\mathrm{R}}\right| $ should exceed $\Omega$ to have the superfluid phase. Since in that case  $\sum_{\alpha}E^{\varphi=0}_{\alpha,\mathbf{k}}$ is independent of $\Omega$, the superfuid ground energy would be the same as in the limit of zero Raman coupling. As in the previous cases, from the gap equations we have $\Delta_{\mathrm{I}}=0$ and thus $\Delta=|\Delta_{\mathrm{R}}|$ for all $\Omega$.
To evaluate a possibility of a metastable state and a realistic  border of phase transitions, we will consider  analytical results for  the effect of the Raman coupling $\Omega$ in two regimes.

\begin{figure}[tbp]
\includegraphics[width=2.6 in]{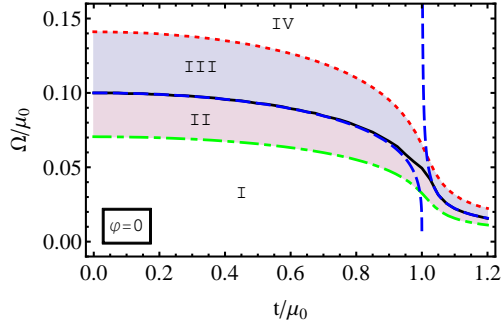}
\caption{ (Color online)
The phase diagram for  the system under the Zeeman field $\Omega$
with a conventional tunneling $t$ for $\protect\varphi=0$.  Below the solid
(Black) line the BCS is formed. In this area the dash-dotted (Green) line  shows a transition from the Regime I
corresponding to the 2D BCS to the Regime II where a metastable normal state is possible.
In the BCS  Regimes I and II, the order parameter doesn't dependent on $\Omega$ and is the
same as in Fig.\protect\ref{fig2}. Above the solid (Black) line there is the metastable superfuild state (Regime III) and the
normal state (Regime IV).}
\label{fig4}
\end{figure}

For small tunnelling regime, $t\leq \mu-\Omega $, the ground energy becomes (see the Section IIC of the Supplementary Material \cite{Junhui2016})
\begin{equation}
\frac{2\pi }{mA}\mathcal{E}=\frac{2\pi }{mA}\mathcal{E}^{\Omega =0}+\Theta
(\Omega -\Delta )g(\Delta,\Omega).  \label{eneergy2}
\end{equation}
Similar to the single layer limit ($t=0$), one goes through four regimes with increasing the Raman coupling, $\Omega$: (I) When $0<\Omega <\sqrt{\epsilon _{b}\mu_t/2}$ with $\mu_t=\sqrt{\mu^{2}-t^{2}}$, the ground state is superfluid with the order parameter being $\Delta =\sqrt{2\epsilon _{b}\mu_t}$. (II) When $\sqrt{\epsilon _{b}\mu_t/2}<\Omega <\sqrt{\epsilon _{b}\mu_t}$, the ground state is still a superfluid phase with $\Delta =\sqrt{2\epsilon _{b}\mu_t}$, but the normal state becomes metastable. (III) When $\sqrt{\epsilon _{b}\mu_t}<\Omega <\sqrt{2\epsilon_{b}\mu_t}$, the ground state becomes a normal state with a metastable superfluid order parameter: $\Delta =\sqrt{2\epsilon _{b}\mu_t}$. (IV) When $\Omega >\sqrt{2\epsilon _{b}\mu_t}$, the superfluid order disappears completely. The four regimes are shown in Fig.\,\ref{fig4}.

In the strong tunneling regime, $t\geq \mu+\Omega $, the ground energy can be expressed to be (see the Section IIC of the Supplementary Material \cite{Junhui2016})
\begin{eqnarray}
\frac{2\pi }{m A}\mathcal{E}&=&\frac{2\pi }{m A}\mathcal{E}^{\Omega
=0}+\frac{1}{2}\Theta (\Omega -\Delta )g(\Delta,\Omega).  \label{eneergy3}
\end{eqnarray}
Similar to previous discussion, the four regimes as a function of Raman coupling, $\Omega$ can be also obtained analytically. Since this does not provide essentially new results, there is no need to present such analytic expression here. However, as one can see in the numerical phase diagram shown in Fig.\,\ref{fig4}, the regimes II and III are shrinking in the large tunneling limit, because the superfluid order parameter is also decreasing. In other words, for $\varphi=0$, the ground state phase diagram is qualitatively similar to the single layer case ($t=0$), because the inter-layer tunneling couples the two layers in the same way for both spin states (without a phase difference).

~\\
\noindent{\bf Raman coupling with $\varphi=\pi/2$}\\
Now we consider the case where $\varphi=\pi/2$, with a finite Raman coupling $\Omega$ and interlayer tunneling $t$. In such a situation the tunneling involves a spin-flip (in the rotated basis). Eigenvalues of the BdG operator now are given by
\begin{equation}  \label{eigensfli}
E^{\pi/2}_{a,\mathbf{k}}=E_{\pm ,\mathbf{k}}\equiv\sqrt{\varepsilon_{\mathbf{%
k}} ^{2}+\Delta ^{2}+t^{2}+\Omega^{2}\pm 2F_{\mathbf{k}}}
\end{equation}
with $F_{\mathbf{k}}=$ $\sqrt{\varepsilon_{\mathbf{k}}^{2}\left(t^{2}+\Omega^{2}\right) +\Delta_{\mathrm{I}}^{2}t^{2}+\Delta ^{2}\Omega^{2}}$. The eigenvalues $E^{\pi/2}_{a,\mathbf{k}}$ are twice degenerate, like the corresponding noninteracting single particle spectrum $\xi^{\pi/2}_{j\gamma,\mathbf{k}}=\varepsilon_\mathbf{k} \pm \sqrt{\Omega^{2}+t^{2}}$.

By having ${\partial \mathcal{E}(\Delta, \Delta_{\mathrm{I}})}/{\partial\Delta_I}=0$, {\ we get $\Delta_{\mathrm{I}}=0$. The gap equation ${\partial \mathcal{E}(\Delta, \Delta_{\mathrm{I}})}/{\partial\Delta}=0$ thus takes the form
\begin{equation} \label{orders}
\sum_{\mathbf{k}} \left[\frac{1+ {\Omega^{2}}/{F_{\mathbf{k}}}}{E_{+,\mathbf{%
k}}}+\frac{1-{\Omega^{2}}/{F_{\mathbf{k}}}}{E_{-, \mathbf{k}}}-\frac{4}{%
{\mathbf{k}}^2/m+\epsilon _{b}}\right]=0.
\end{equation}

\begin{figure}[tbp]
\includegraphics[width=2.6 in]{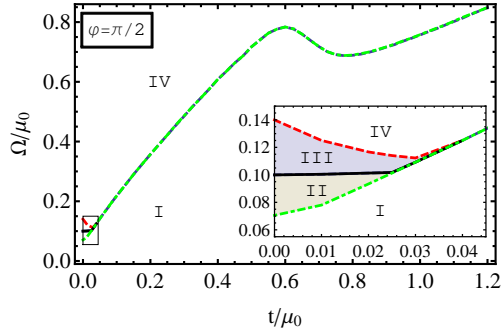}
\caption{ (Color online) Phase diagram  in the $t$-$\Omega$ plane for $\protect\varphi=\protect\pi/2$.
The solid
(Black) line represents  a boundary for the  first order phase transition determined by
minimizing  the energy. In the Regime II the normal
state becomes metastable and in the Regime III
the superfluid state becomes a metastable state.  Close to the origin the phase diagram
is magnified in the insert. }
\label{fig5}
\end{figure}
In Fig.\,\ref{fig5}, we show the phase diagram in terms of the tunneling $t$ and the Raman coupling $\Omega$. In the range of small $t$ displayed in the insert of Fig.\,\ref{fig5}, there are four Regimes I-IV, as in the previously considered cases. However, when the tunneling amplitude becomes larger,  the range of superfluid phase increases significantly. This is very different from the phase diagram for $\varphi=0$  shown in Fig. \ref{fig4}. Therefore a much stronger Raman coupling (effective magnetic field) is now required to destroy the superfluid phase (Regime I) which now goes directly to the normal phase (Regime IV) without passing the metastable phases (Regimes II and III). Such a phase transition is of the second  order, a feature absent in the previously considered cases where the phase transition is of the  first order. The first order phase transition now occurs only for small tunneling ($t<0.04\mu_0$)  where the superfluid state (Regime I)  first goes to the metastable states (Regimes II and III) before reaching the normal state (Regime IV).

Note that the nature of the phase transition between Regime I (superfluid) and Regime IV (normal) is determined by the  meta stable solutions in-between them, i.e., Regimes II and III, in the small $t$ limit. When the interlayer tunneling is stronger than $0.04 \mu_0$, the intermediate regime disappears because opposite spins in different layers are mixed due to the spin-flip inter-layer tunneling (Fig.\,1(b)), making the superfluid state in the s-wave pairing channel hardly to form in Regimes II and III. As a result, the phase transition for large $t$ becomes fully determined by the curvature of free energy ($\partial ^{2}\mathcal{E}/\partial \Delta ^{2}$) at $\Delta=0$, the same as the condition determining the boundary between Regimes I and II.

\begin{figure}[tbp]
\includegraphics[width=2.6 in]{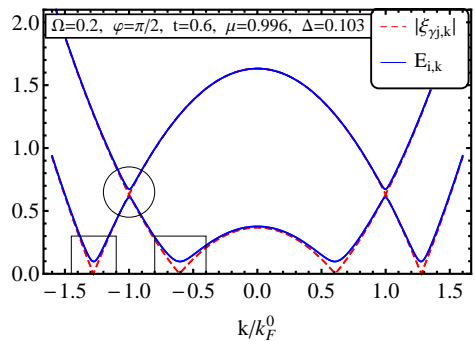}
\caption{(Color online) The excitation spectrum for $\protect\varphi=\protect\pi/2,
\Omega=0.2\protect\mu_0$ and $t=0.6\protect\mu_0$.
For a finite tunneling strength $t$, gaps shown in Rectangles open at the Fermi surface through the triplet pairing. On the other hand, when $\Omega$ becomes finite, a gap starts opening above the Fermi surface (shown in the Circle).
}
\label{fig6}
\end{figure}

This result is very unusual.  Normally  the effective magnetic field $\Omega$ and tunneling $t$ reduce the superfluid properties by breaking the Cooper pairs through the Zeeman effects. As we can see from the single particle eigenenergies, the non-interacting part of the Hamiltonian $\mathbf{{\Xi}_k^{\pi/2}} = \varepsilon_\mathbf{k} + \sqrt{t^2+\Omega^2}\,\sigma _{z}$ shows an even larger effective Zeeman field $\sqrt{t^2+\Omega^2}$ when including both effects, the Raman coupling and interlayer tunneling. The magnetism can be defined as $M=\frac{N_\uparrow-N_\downarrow}{N_\uparrow+N_\uparrow}$, where $N_\gamma=\sum_{j,\bfk}\langle c_{j\gamma,\bfk}^\dag c_{j\gamma,\bfk}\rangle$ is a number of atoms with a spin $\gamma$. Thus in the limit of weak Raman and interlayer coupling, $\mu>\sqrt{t^2+\Omega^2}$,  the magnetisation is $M=-\sqrt{t^2+\Omega^2}/{\mu}$, while for $\mu\leq\sqrt{t^2+\Omega^2}$, the system  is  fully magnetized,  $M=-1$. In other words, larger $\Omega$ and $t$ mean a larger magnetism.

However, Fig.\,\ref{fig5}  indicates that for larger values of $t$ and $\Omega$ their influence is mutually canceled out, so that the effect of the  magnetic field becomes much smaller and correspondingly the superfluid region is broadened. This is due to  a specific form of atom-atom interaction in the bilayer system. The interaction is now represented by the  term $D_{\pi/2}$ entering the BdG operator:
\begin{eqnarray}
D_{\pi/2}&=&-i \Delta \sin\theta \big[(c^\dag_{\Uparrow\uparrow,
\mathbf{k}}c^\dag_{\Downarrow\uparrow,-\mathbf{k}}+ c^ \dag_{\Uparrow\downarrow,\mathbf{k}%
}c^ \dag_{\Downarrow\downarrow,-\mathbf{k}})-(\Uparrow\leftrightarrow \Downarrow )\big]\notag \\&&
+ \Delta\cos\theta\big[(c^
\dag_{\Uparrow\uparrow,\mathbf{k}}c^ \dag_{\Uparrow\downarrow,-\mathbf{k}}+ c^
\dag_{\Downarrow\uparrow,\mathbf{k}}c^ \dag_{\Downarrow\downarrow,-\mathbf{k}})-(\uparrow\leftrightarrow \downarrow )\big], \label{dpha}
\end{eqnarray}
where  $\tan\theta\equiv t/\Omega$ and we have used the fact that $\Delta_{\mathrm{I}}=0 $. The first term in  $D_{\pi/2}$ indicates that a triplet pairing forms for atoms residing at different layers if the interlayer tunneling $t$ is sufficiently large. This opens a gap in the excitation spectrum at the Fermi surface,  as one can see in Fig.\ref{fig6}. The second term represents a singlet pairing within the same layer. This opens two additional gaps at higher energies above the Fermi surface (see Fig.\ref{fig6}). The ratio $\tan\theta\equiv t/\Omega$ measures the relative strength between the singlet and the
triplet pairing in the spin space.  Increasing spin-flip tunneling enhances the interlayer spin triplet pairing and thus  makes the large Zeeman field $\Omega$ to loose its efficiency in destroying the Cooper pairs.

Finally, we explore a situation where $0<\varphi<\pi/2$, so that both  $\Omega_x=\Omega\cos\varphi$ and $\Omega_y=\Omega\sin\varphi$ are non-zero. In Fig.\,\ref{fig7}, we show the phase diagram for three different finite values of $\Omega_x$, i.e., for different magnitudes of the parallel Zeeman field.
Although the increase in the superfluid pairing is still significant, the extent of the superfluid regime reduces for larger $\Omega_x$.  In the rotated basis $\Psi _{\mathbf{k}}$, an increase of $\Omega_x$ enhances the importance of the conventional tunneling with respect to the spin-flip tunneling. The conventional tunnelling determined by $\Omega_x$  has a tendency to destroy the degenerate structure in the spectrum, so that it prevents formation of triplet pairing and reduces the superfluidity, unlike the spin-flip tunneling which is determined by $\Omega_y$. Furthermore, the phase boundary due to the co-existence of
a meta-stable state becomes much broader, compared to the case with zero $\Omega_x$. In Fig.\ref{fig8}, we show the order parameter with respect to $t$ for a
fixed $\Omega_y=0.2\mu_0$.  When $t=0$ and $\Omega_x=0$, the Cooper pair is a complete spin singlet,  so the finite Raman coupling of $\Omega_y$ prevents
formation of Cooper pairs.  Increasing the tunneling amplitude ($t$) enhances  the triplet pairing.  Thus for a sufficient large $t$, the system  undergoes a transition to the superfluid from  the normal state. For  finite $\Omega_x$ and $\Omega_y=0.2\mu_0$,  there is a conventional tunneling  in addition to the spin-flip tunneling in the rotated basis. In that case  the superfluid  formes in a narrow range of tunneling  values $t$.
~\\

\begin{figure}[tbp]
\includegraphics[width=2.6 in]{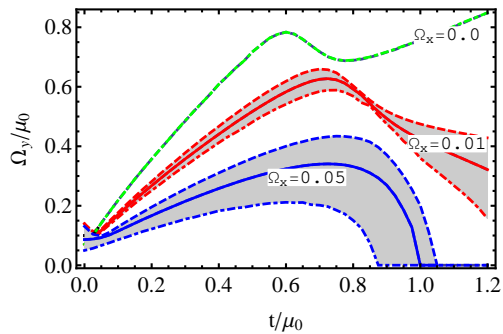}
\caption{(Color online) Phase diagram in the  $t$-$\Omega_y$ plane for a finite value of $%
\Omega_x$. The solid line is determined by minimizing energy and in the
shaded area, there metastable states  exist.}
\label{fig7}
\end{figure}

\begin{figure}[htb]
\centering
\vspace{5 mm}
\includegraphics[width=2.6 in]{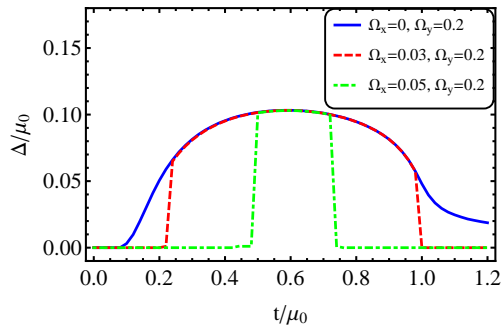}
\caption{(Color online) The order parameter vs. the tunneling strength for $%
\Omega_x= (0,~0.03,~0.05)\protect\mu_0$ and $\Omega_y=0.2\protect\mu_0$. The
lines are determined by minimizing the energy. The metastable
state regime is not shown here.}
\label{fig8}
\end{figure}

~\\
\noindent{\bf \large Conclusions}\\
\noindent  We have explored a new mechanism to greatly enhance superfluidity of ultracold Fermi gases in a large range of the effective magnetic field.
The mechanism can be implemented for a bilayer atomic system subjected to an interlayer tunneling. Additionally a Raman coupling induces intralayer spin-flip transitions with a phase difference between the two layers. Such a Raman coupling serves as a magnetic field staggered in different layers.  After introducing a proper gauge transformation, one arrives at a non-staggered magnetic field and a spin-flip tunnelling between the layers.  In such a situation the Cooper pairs  were shown to acquire a component due to the triplet pairing. This supports a co-existence of the superfluidity for a much stronger effective magnetism.
Our findings are helpful for understanding and controlling the superconductivity in the presence of the magnetic fields.

~\\
\noindent{\bf \large Acknowledgements}\\
\noindent We thank C.-Y. Mou, B. Xiong, J.-S. You and E. Demler for useful discussion. This work is supported by NCTS and MoST grant in Taiwan.

~\\
\noindent{\bf \large Author Contributions}\\
D.-W.W. and G. J. conceived the idea. J.-H.Z. has carried out the analytical and the numeric calculations. D.-W.W. and G.J. have also contributed to the analytical calculations. D.-W.W., J.-H.Z. and G.J. discussed results and prepared the manuscript.

\newpage

\begin{center}
  { \Large \bf Supplementary materials for `Superfluidity enhanced by spin-flip tunnelling
in the presence of a magnetic field'}
\end{center}

\section{Derivation of $D_{\varphi}$}

According to Eq.(10) of the main text, applying the BCS
mean-field approximation, the interaction term takes the form
\begin{equation}
H_{\mathrm{int}}=\frac{2\Delta^{2}A}{\left\vert g\right\vert }+\sum_{\mathbf{k,}j=\Uparrow,\Downarrow}\left(\Delta_{j}\psi_{j\uparrow,\mathbf{k}}^{\dagger}\psi_{j\downarrow,-\mathbf{k}}^{\dagger}+H.c.\right),\label{hint-mf-k}
\end{equation}
where $\Delta=\left\vert \Delta_{\Uparrow}\right\vert =\left\vert \Delta_{\Downarrow}\right\vert $
, with $\Delta_{\Uparrow}^{\ast}=\Delta_{\Downarrow}=\Delta_{\mathrm{R}}+i\Delta_{\mathrm{I}}$.
By using the Fermi communtation relations, the second term of Eq.(\ref{hint-mf-k}),
denoted as $V_{\mathrm{int}}$, can be represented in a symmetrised
manner as
\begin{equation}
V_{\mathrm{int}}\equiv\frac{1}{2}\sum_{\mathbf{k}j}\left[\Delta_{j}(\psi_{j\uparrow,\mathbf{k}}^{\dagger}\psi_{j\downarrow,-\mathbf{k}}^{\dagger}-\psi_{j\downarrow,\mathbf{k}}^{\dagger}\psi_{j\uparrow,-\mathbf{k}}^{\dagger})+H.c.\right]\,.\label{hint-mf-k-initial}
\end{equation}
By defining a row field operator $\Psi_{\mathbf{k}}^{\dagger}=[\psi_{\Uparrow\uparrow,\mathbf{k}}^{\dagger},\psi_{\Uparrow\downarrow,\mathbf{k}}^{\dagger},\psi_{\Downarrow\uparrow,\mathbf{k}}^{\dagger},\psi_{\Downarrow\downarrow,\mathbf{k}}^{\dagger}]$,
we obtain a concise form
\begin{equation}
V_{\mathrm{int}}\equiv\frac{1}{2}\sum_{\mathbf{k}}\Psi_{\mathbf{k}}^{\dagger}\left(i\Delta_{\mathrm{R}}\tau_{0}\otimes\sigma_{y}+\Delta_{\mathrm{I}}\tau_{z}\otimes\sigma_{y}\right)\,\Psi_{-\mathbf{k}}^{\dagger T}+H.c.,\label{eq:H_int-orig-basis}
\end{equation}
where $^{T}$ stands for a transposed matrix, so $\Psi_{-\mathbf{k}}^{\dagger T}$
is a column field operator. In a diagonal representation
of the single particle problem, the field operators read
\begin{equation}
C_{\mathbf{k}}=V_{\varphi}\Psi_{\mathbf{k}}\,,\qquad C_{\mathbf{k}}^{\dagger}=\Psi_{\mathbf{k}}^{\dagger}V_{\varphi}^{\dagger}=\Psi_{\mathbf{k}}^{\dagger}V_{\varphi}^{-1}\,,\label{eq:c_k}
\end{equation}
where $\Psi_{\mathbf{k}}\equiv\left(\Psi_{\mathbf{k}}^{\dagger}\right)^{\dagger}=[\psi_{\Uparrow\uparrow,\mathbf{k}},\psi_{\Uparrow\downarrow,\mathbf{k}},\psi_{\Downarrow\uparrow,\mathbf{k}},\psi_{\Downarrow\downarrow,\mathbf{k}}]^{T}$
is a column matrix. Therefore the field operators entering Eq.(\ref{eq:H_int-orig-basis})
can be represented as in terms of the normal modes as
\begin{equation}
\Psi_{\mathbf{k}}^{\dagger}=C_{\mathbf{k}}^{\dagger}V_{\varphi}\,,\qquad\Psi_{\mathbf{k}}^{\dagger T}=V_{\varphi}^{T}C_{\mathbf{k}}^{\dagger T}\,,\label{eq:psi_k-Suppl}
\end{equation}
giving
\begin{equation}
V_{\mathrm{int}}\equiv\frac{1}{2}\sum_{\mathbf{k}}\left(C_{\mathbf{k}}^{\dagger}D_{\varphi}\,C_{-\mathbf{k}}^{\dagger T}+C_{-\mathbf{k}}^{T}D_{\varphi}^{\dagger}\,C_{\mathbf{k}}\right)\,,\label{eq:V_int}
\end{equation}
 with
\begin{equation}
D_{\varphi}=V_{\varphi}\left(i\Delta_{\mathrm{R}}\tau_{0}\otimes\sigma_{y}+\Delta_{\mathrm{I}}\tau_{z}\otimes\sigma_{y}\right)V_{\varphi}^{T}\,.\label{eq:D_varphi-Suppl}
\end{equation}
where we used the fact that the unitary transformation
$V_{\varphi}$ is independent of $\mathbf{k}$. The operator $V_{\mathrm{int}}$
can be rewritten in a matrix form as
\begin{equation}
V_{\mathrm{int}}=\frac{1}{2}\sum_{\mathbf{k}}[C_{\mathbf{k}}^{\dagger},C_{\mathbf{-k}}^{T}]\left[\begin{array}{cc}
0 & D_{\varphi}\\
D_{\varphi}^{\dagger} & 0
\end{array}\right]\left[\begin{array}{c}
C_{\mathbf{k}}\\
C_{\mathbf{-k}}^{\dagger T}
\end{array}\right],\label{V-int-Suppl}
\end{equation}
giving the interaction term featured in Eq.(11) of the main text.

Specifically for $\varphi=0$, we have $V_{0}=\exp[-i\frac{\pi}{4}\tau_{y}]$
and $V_{0}^{T}=\exp[i\frac{\pi}{4}\tau_{y}]$, which satisfies $V_{0}\tau_{z}V_{0}^{T}=\tau_{x}$,
giving
\begin{equation}
D_{0}=i\Delta_{\mathrm{R}}\tau_{0}\otimes\sigma_{y}+\Delta_{\mathrm{I}}\tau_{x}\otimes\sigma_{y}\,.
\end{equation}
For another case $\varphi=\pi/2$, we have $V_{\pi/2}=\exp[-i\frac{\theta}{2}\tau_{y}\otimes\sigma_{y}]$
and $V_{\pi/2}^{T}=V_{\pi/2}$. Therefore
\begin{eqnarray}
V_{\pi/2}\tau_{0}\otimes\sigma_{y}V_{\pi/2} & = & \left(\tau_{0}\otimes\sigma_{y}\cos\theta-i\tau_{y}\otimes\sigma_{0}\sin\theta\right)\,,\\
V_{\pi/2}\tau_{z}\otimes\sigma_{y}V_{\pi/2} & = & \tau_{z}\otimes\sigma_{y},
\end{eqnarray}
giving
\begin{equation}
D_{\pi/2}=\Delta_{\mathrm{R}}\left(i\tau_{0}\otimes\sigma_{y}\cos\theta+\tau_{y}\otimes\sigma_{0}\sin\theta\right)+\Delta_{\mathrm{I}}\tau_{z}\otimes\sigma_{y}.
\end{equation}

\section{Ground state energy}

Let us denote the eigenvalues of the Bogoliubov-DeGuinne
term to be $E_{\alpha,\mathbf{k}}^{\varphi}\geq0$ with $\alpha=1,2,3,4$
and their corresponding operators as $d_{\alpha\bf{k}}$. The
Hamiltonian in this eigenbasis reads:
\begin{eqnarray}
H & = & \frac{1}{2}\sum_{\alpha, \mathbf{k}}\left\{ E_{\alpha,\mathbf{k}}^{\varphi}d_{\alpha,\bf{k}}^{\dag}d_{\alpha,\bf{k}}-E_{\alpha,\mathbf{k}}^{\varphi}d_{\alpha,\bf{-k}}d_{\alpha,\bf{-k}}^{\dag}\right\} +2\sum_{\mathbf{k}}\varepsilon_{\mathbf{k}}+\frac{2\Delta^{2}A}{\left\vert g\right\vert }\notag\\
 & = & \sum_{\alpha,\mathbf{k}}E_{\alpha,\mathbf{k}}^{\varphi}d_{\alpha,\bf{k}}^{\dag}d_{\alpha,\bf{k}}-\frac{1}{2}\sum_{\alpha,\mathbf{k}}E_{\alpha,\mathbf{k}}^{\varphi}+2\sum_{\mathbf{k}}\varepsilon_{\mathbf{k}}+\frac{2\Delta^{2}A}{\left\vert g\right\vert }.
\end{eqnarray}
At zero temperature, the ground state has a zero number
of Bololiubov excitations $\left\langle d_{\alpha,\bf{k}}^{\dag}d_{\alpha,\bf{k}}\right\rangle =0$
because all $E_{\alpha,\mathbf{k}}^{\varphi}\geq0$. Thus the ground-state
energy becomes
\begin{equation}
\mathcal{E}=\frac{2\Delta^{2}A}{\left\vert g\right\vert }+\sum_{\mathbf{k}}\left(2\varepsilon_{\mathbf{k}}-\frac{1}{2}\sum_{\alpha}E_{\alpha,\mathbf{k}}^{^{\varphi}}\right).\label{k0groundenergy}
\end{equation}

\subsection{Ground energy for single layer limit}

In the following, we give the calculation for the ground energy for
the case $t=0$, so that $E_{\alpha,\mathbf{k}}^{\varphi}=|\sqrt{\left(\frac{k^{2}}{2}-\mu\right)^{2}+\Delta^{2}}\pm\Omega|$.
Explicitly, the ground energy becomes
\begin{equation}
\mathcal{E}=2\Delta^{2}\sum_{\mathbf{k}}\frac{1}{\mathbf{k}^{2}/m+\epsilon_{b}}+\sum_{\mathbf{k}}(\frac{\mathbf{k}^{2}}{m}-2\mu)-\sum_{\mathbf{k},\pm}\left\vert \sqrt{\left(\frac{\mathbf{k}^{2}}{2m}-\mu\right)^{2}+\Delta^{2}}\pm\Omega\right\vert .
\end{equation}
Note that $\sum_{\mathbf{k}}=\frac{1}{\left(2\pi\right)^{2}}A\int d^{2}\mathbf{k}=\frac{mA}{2\pi}\int\frac{k}{m}dk$,
where $A$ is the area of the system. We use the replacement$\frac{k}{\sqrt{m}}\rightarrow k$,
thus we obtain
\begin{equation}
\frac{2\pi\mathcal{E}}{mA}=2\Delta^{2}\int kdk\frac{1}{k^{2}+\epsilon_{b}}+\int k^{3}dk-\int2\mu kdk-\sum_{\pm}\int kdk\left\vert \sqrt{\left(\frac{k^{2}}{2}-\mu\right)^{2}+\Delta^{2}}\pm\Omega\right\vert .\label{energys}
\end{equation}
We will set a cutoff momentum $\lambda$ for the integral,
and then expand the result by $1/\lambda$. Finally we let the cutoff
$\lambda$ to be infinite. For the first term in the Eq.(\ref{energys}),
we have
\begin{eqnarray}
2\Delta^{2}\int_{0}^{\lambda}kdk\frac{1}{k^{2}+\epsilon_{b}} & = & \Delta^{2}\ln\left[\lambda^{2}+\epsilon_{b}\right]-\Delta^{2}\ln\epsilon_{b}\notag\\
 & = & \Delta^{2}\ln\lambda^{2}-\Delta^{2}\ln\epsilon_{b}+\Delta^{2}\cdot O(\frac{\epsilon_{b}}{\lambda^{2}}).
\end{eqnarray}
On the other hand, the second and third terms are
\begin{equation}
\int_{0}^{\lambda}k^{3}dk=\frac{1}{4}\lambda^{4}
\end{equation}
and
\begin{equation}
\int_{0}^{\lambda}2\mu kdk=\mu\lambda^{2}
\end{equation}
respectively.

For the last term in Eq.(\ref{energys}), we consider two different
cases. For $\Delta>\Omega$, we have $\sqrt{\left(\frac{k^{2}}{2}-\mu\right)^{2}+\Delta^{2}}\pm\Omega>0$.
In that case the integral becomes
\begin{eqnarray}
a_{4} & \equiv & \sum_{\pm}\int_{0}^{\lambda}kdk\left\vert \sqrt{\left(\frac{k^{2}}{2}-\mu\right)^{2}+\Delta^{2}}\pm\Omega\right\vert \notag\\
 & = & \frac{1}{2}\int_{0}^{\lambda^{2}}dk^{2}\sqrt{\left(k^{2}-2\mu\right)^{2}+4\Delta^{2}}\notag\\
 & = & \frac{1}{2}\int_{-2\mu}^{\lambda^{2}-2\mu}dx\sqrt{x^{2}+4\Delta^{2}}.
\end{eqnarray}
By using
\begin{equation}
\int dx\sqrt{x^{2}+y^{2}}=\frac{1}{2}x\sqrt{x^{2}+y^{2}}+\frac{1}{2}y^{2}\ln\left\vert \frac{\sqrt{x^{2}+y^{2}}+x}{y}\right\vert +Const.
\end{equation}
we have
\begin{eqnarray}
a_{4} & = & \left[\frac{1}{4}\left(\lambda^{2}-2\mu\right)^{2}+\frac{\Delta^{2}}{2}+\Delta^{2}O\left(\frac{\Delta^{2}}{\lambda^{4}}\right)\right]+\Delta^{2}\left[\ln\frac{\lambda^{2}-2\mu}{2\Delta^{2}}+O\left(\frac{\Delta^{2}}{\lambda^{4}}\right)\right]\notag\\
 &  & +\mu\sqrt{\mu^{2}+\Delta^{2}}-\Delta^{2}\ln\frac{\sqrt{\mu^{2}+\Delta^{2}}-\mu}{2\Delta^{2}},
\end{eqnarray}
As a result, for $\Delta>\Omega$ and $\lambda\rightarrow\infty$,
we get
\begin{equation}
\frac{2\pi\mathcal{E}}{mA}=\Delta^{2}\ln\frac{\sqrt{\mu^{2}+\Delta^{2}}-\mu}{\epsilon_{b}}-\frac{\Delta^{2}}{2}-\mu\sqrt{\mu^{2}+\Delta^{2}}-\mu^{2}\equiv f\left(\Delta,\epsilon_{b},\mu\right).
\end{equation}

For $\Delta<\Omega,$  one has
\begin{eqnarray}
 &  & -\sum_{\pm}\int kdk\left\vert \sqrt{\left(\frac{k^{2}}{2}-\mu\right)^{2}+\Delta^{2}}\pm\Omega\right\vert \notag\\
 & = & -2\int kdk\sqrt{\left(\frac{k^{2}}{2}-\mu\right)^{2}+\Delta^{2}}+g\left(\Delta,\Omega,\mu\right),\label{theta}
\end{eqnarray}
where
\begin{equation}\label{region}
  g\left(\Delta,\Omega,\mu\right)=2\int_{S}kdk[\sqrt{\left({k^{2}}/{2}-\mu\right)^{2}+\Delta^{2}}-\Omega]
\end{equation}
and $S$ is the region that $\sqrt{\left(\frac{k^{2}}{2}-\mu\right)^{2}+\Delta^{2}}<\Omega,$
i.e., $\mu-\sqrt{\Omega^{2}-\Delta^{2}}<\frac{k^{2}}{2}<\mu+\sqrt{\Omega^{2}-\Delta^{2}}.$
In the following, we suppose that $\mu-\sqrt{\Omega^{2}-\Delta^{2}}>0$
and try to calculate out the integral $g\left(\Delta,\Omega,\mu\right)$.
(Note that if $\mu+\sqrt{\Omega^{2}-\Delta^{2}}<0$, then the region
$S$ vanishes.) Performing the integration, one has
\begin{equation}
2\int_{S}kdk\sqrt{\left(\frac{k^{2}}{2}-\mu\right)^{2}+\Delta^{2}}=2\Omega\sqrt{\Omega^{2}-\Delta^{2}}+\Delta^{2}\ln\frac{\Omega+\sqrt{\Omega^{2}-\Delta^{2}}}{\Omega-\sqrt{\Omega^{2}-\Delta^{2}}},\label{theta1}
\end{equation}
and
\begin{equation}
-2\int_{S}kdk\Omega=-\int_{2\left[\mu-\sqrt{\Omega^{2}-\Delta^{2}}\right]}^{2\left[\mu+\sqrt{\Omega^{2}-\Delta^{2}}\right]}dk^{2}\Omega=-4\Omega\sqrt{\Omega^{2}-\Delta^{2}}.\label{theta2}
\end{equation}
Note that these two integrals are independent on $\mu$, so
one can write $g\left(\Delta,\Omega\right)\equiv g\left(\Delta,\Omega,\mu\right)$.
Finally, we have
\begin{equation}
\frac{2\pi\mathcal{E}}{mA}=f\left(\Delta,\epsilon_{b},\mu\right)+\Theta\left(\Omega-\Delta\right)g\left(\Delta,\Omega\right),
\end{equation}
where $g\left(\Delta,\Omega\right)=\Delta^{2}\ln\frac{\Omega+\sqrt{\Omega^{2}-\Delta^{2}}}{\Omega-\sqrt{\Omega^{2}-\Delta^{2}}}-2\Omega\sqrt{\Omega^{2}-\Delta^{2}}$.

\subsection{Ground energy for a zero Raman coupling limit}

In the limit of  zero Raman coupling limit,
one has $E_{\alpha,\mathbf{k}}^{\Omega=0}\equiv\sqrt{\left(\frac{k^{2}}{2m}-\mu\pm t\right)^{2}+\Delta^{2}}$
, where we have used the fact $\Delta_{\mathrm{I}}=0$.
It has two effective chemical potentials $\mu_{\pm}=\mu\pm t$.  Using the result in the last
subsection, it is easy to obtain the ground energy
\begin{equation}
\frac{2\pi\mathcal{E}^{\Omega=0}}{mA}=\frac{1}{2}f\left(\Delta,\epsilon_{b},\mu+t\right)+\frac{1}{2}f\left(\Delta,\epsilon_{b},\mu-t\right).
\end{equation}

\subsection{Ground energy Raman coupling with $\varphi=0$}

For $\varphi=0$, $\ $we have $E_{\alpha,\mathbf{k}}^{0}=\left\vert \sqrt{\left(\frac{k^{2}}{2m}-\mu_{\pm}\right)^{2}+\Delta^{2}}\pm\Omega\right\vert $,
where $\mu_{\pm}=\mu\pm t$ and we have used the fact ${\Delta_{\mathrm{I}}=0}$.
Similar to the previous cases, for $\Delta>\Omega,$
we have ground energy
\begin{equation}
\frac{2\pi\mathcal{E}}{mA}=\frac{2\pi\mathcal{E}^{\Omega=0}}{mA}.
\end{equation}
On the other hand, for $\Delta<\Omega$ and $\mu_{\pm}-\sqrt{\Omega^{2}-\Delta^{2}}>0$
, using the method in Eqs. (\ref{theta}), (\ref{theta1}) and (\ref{theta2}),
we have
\begin{equation}
\frac{2\pi}{mA}\mathcal{E}=\frac{2\pi}{mA}\mathcal{E}^{\Omega=0}+\Theta(\Omega-\Delta)g\left(\Delta,\Omega\right).
\end{equation}
For another case,  $\Delta<\Omega$, $\mu_{-}+\sqrt{\Omega^{2}-\Delta^{2}}<0$
and $\mu_{+}-\sqrt{\Omega^{2}-\Delta^{2}}>0$, where the integral region $S$
vanishes (See Eq.(\ref{region})) for the $\mu_{-}$ branch, the ground energy
becomes
\begin{equation}
\frac{2\pi}{mA}\mathcal{E}=\frac{2\pi}{mA}\mathcal{E}^{\Omega=0}+\frac{1}{2}\Theta(\Omega-\Delta)g\left(\Delta,\Omega\right).
\end{equation}
\end{document}